\documentclass[%
 preprint, 
 amsmath,amssymb,
 aps, physrev,
]{revtex4-2}

\usepackage{graphicx}
\usepackage{dcolumn}
\usepackage{bm}
\usepackage{soul} 
\usepackage{xcolor}

\usepackage[normalem]{ulem}

\begin{document}


\title{
\textbf{
First-Principles Prediction of Phonon-Mediated Infrared Optical Properties of WO$_3$ Polymorphs
}}

\author{Sreerag Sundaram}
\affiliation{Department of Energy Science and Engineering, Indian Institute of Technology Bombay, Powai, Mumbai - 400076, India}
 

\author{Karthik Sasihithlu}
\email{ksasihithlu@ese.iitb.ac.in}
\affiliation{Department of Energy Science and Engineering, Indian Institute of Technology Bombay, Powai, Mumbai - 400076, India.}


\date{\today}

\begin{abstract}
Many crystals exhibit polymorphism, undergoing atomic rearrangements that result in unit cells belonging to different symmetry groups. These structural changes directly affect lattice vibration modes and consequently influence their mid-infrared optical properties. In this study, we use tungsten trioxide (WO$_3$) as a representative system, owing to its multiple temperature-dependent phases, to study the impact of polymorphism on mid-infrared optical behaviour. Using a fully first-principles approach, we investigate three phases of WO$_3$ and evaluate their mid-infrared optical 
properties. Significant differences are observed among the three crystallographic phases, demonstrating the potential of this methodology as a predictive tool for materials discovery and targeted design.
\end{abstract}

\maketitle


\section{\label{sec:Intro} Introduction}


The properties of crystalline materials are determined by their constituent 
atomic species and their arrangement within the crystal lattice. The 
occupation of lattice sites governs the thermodynamic stability under given 
conditions such as pressure and temperature. Crystal symmetry governs electronic structure, lattice dynamics, 
and the resulting thermal and optical properties.
In particular, crystal symmetry determines the allowed 
vibrational modes and their degeneracies, thereby governing phonon-mediated 
interactions with electromagnetic radiation. This connection is especially 
important in the infrared regime, where optical response arises primarily 
from lattice vibrations.

In this work, we focus on polymorphic materials, i.e., systems with identical chemical composition that exist in multiple crystallographic phases. Each polymorphic phase corresponds to a distinct symmetry group and is stable over a specific range of external conditions.
These differences result in marked variations in material properties. Notable instances of polymorphic materials that have been previously investigated include Ga$_2$O$_3$ \cite{polymorph-Ga2O3, Ga2O3-applications}, TiO$_2$ \cite{polymorph-TiO2, TiO2-applications_1, TiO2-applications_2}, SiO$_2$ \cite{polymorph-SiO2, SiO2-applications_1, SiO2-applications_2}, SiC \cite{polymorph-SiC, SiC-applications_1, SiC-applications_2}, and 2D materials such as BN \cite{polymorph-BN, BN-applications} and MoTe$_2$ \cite{polymorph-MoTe2, MoTe2-applications}, whose polymorphs exhibit distinct optical, electronic, and vibrational characteristics finding use in various applications such as optoelectronics, catalysis, sensors, etc. \cite{Apps-1,Apps-2,Apps-3,Apps-4}.
Phase transformations between polymorphs can be driven by external stimuli such as pressure, temperature, or electric potential \cite{P-based_PhaseTrans, PT-based_PhaseTrans, T-based_PhaseTrans1, T-based_PhaseTrans2, T-based_PhaseTrans3, Volt-based_PhaseTrans1, Volt-based_PhaseTrans2}. 
Among these, we choose a material that exists in multiple crystallographic phases, with structural transformations triggered by temperature and supported by substantial experimental evidence -- tungsten trioxide (WO$_3$) \cite{WO3_polymorphism_1, WO3_polymorphism_2}. 
The most commonly reported 
phases of WO$_3$ are summarized in Table~\ref{tab:Transit_temp_WO3}. Each 
phase remains stable within a specific temperature range and undergoes 
temperature-driven phase transformations. Because $\delta$-WO$_3$ is stable only over a narrow temperature interval and
$\epsilon$-WO$_3$ exists only at temperatures well below room temperature, both phases are excluded from the present study.

\begin{table}[t]

    \caption[Most commonly reported crystal structures of WO$_3$ and their stable temperature ranges]{\label{tab:Transit_temp_WO3} Most commonly reported crystal structures of  WO$_3$ and their stable temperature ranges. The temperatures chosen for further simulations for each polymorph is listed in the last column.}
    
    \begin{ruledtabular}
    \centering
    
    \begin{tabular}{cccc}
        Phase & Crystal Structure & Stable Temperature Range (K) & Chosen Temperature (K) \\
        \colrule
        $\alpha$-WO$_3$ & Tetragonal & $>$ 1020 & 1300 \\
        $\beta$-WO$_3$ & Orthorhombic & 600--1000  & 700 \\
        $\gamma$-WO$_3$ & Monoclinic & 290--600  & 400 \\
        $\delta$-WO$_3$ & Triclinic & 223--290  & -- \\
        $\epsilon$-WO$_3$ & Monoclinic & $<$ 223  & -- \\
    \end{tabular}
    \end{ruledtabular}
\end{table}

Numerous experimental and computational studies have attempted to characterise these phases \cite{WO3_transition_expt_1999, WO3_transition_diffraction_2002, WO3_transition_exptRaman_2002, WO3_transition_expt_2006, WO3_Leige_thesis_2017, WO3_elec_phonon_RI_2019, WO3_transition_2022}. In particular, several first principles investigations have focused on the electronic structure  of WO$_3$, which governs its optical response in the UV--visible--NIR 
region \cite{WO3_DFT_Elec_2011, WO3_DFT_Elec_2013, 
WO3_DFT_Elec_2013_2, WO3_DFT_Elec_2015, WO3_DFT_Elec_2019, 
WO3_DFT_Elec_2025, WO3_DFT_Elec_Phonon_2016}, as well as on its lattice 
dynamics \cite{WO3_Phonon_2016, WO3_DFT_Elec_Phonon_2016, 
WO3_Leige_thesis_2017, WO3_elec_phonon_RI_2019, WO3_Phonon_2022}. 
Refs. \onlinecite{WO3_elec_phonon_RI_2019, WO3_Phonon_2022} analysed the phonon spectra to identify infrared-active modes and calculate IR reflectivity, although details of the IR reflectivity formalism was not described in detail. Moreover, the omission of damping as well as renormalisation effects due to temperature and crystal anharmonicities limit the accuracy of the predicted optical responses. Therefore, the prediction of mid-infrared optical properties from first principles remains relatively unexplored, prompting us to revisit these polymorphs using a more comprehensive framework. 
In this work, we focus exclusively on lattice dynamics and their role in determining the  mid-infrared optical response. The mechanisms responsible for the crystallographic phase transitions themselves are not investigated; instead, we take the experimentally established phase behaviour of WO$_3$ as a given and analyse each phase within its stable structural framework.


A recurring issue in prior studies \cite{WO3_Phonon_2016, WO3_DFT_Elec_Phonon_2016, WO3_elec_phonon_RI_2019, WO3_Phonon_2022} is the presence of imaginary (negative) phonon frequencies in several phases of WO$_3$, likely  arising from the omission of temperature-dependent renormalization effects. Furthermore, to the best of our knowledge, no study has  systematically extracted the complete dielectric function for the polymorphic phases of this material. In this work, we address these gaps by performing a comprehensive phonon analysis and evaluating the mid-infrared optical properties, along with thermal conductivity, of three phases of WO$_3$. We further 
demonstrate that temperature-driven structural phase transitions in such polymorphic systems give rise to thermochromic behaviour.


\section{\label{sec:Computational_details} Computational Details}

All crystal structures of WO$_3$ considered in this work were obtained from the Materials Project database \cite{MatPro_1,MatPro_2}. The selected polymorphs crystallize in the following space groups: monoclinic (P2$_1$/c, No. 14), orthorhombic (Pbcn, No. 60), and tetragonal (P4/nmm, No. 129). The initial structural models were taken directly from the database and subsequently used as inputs for first-principles calculations. All structures were further optimized within the framework of density functional theory (DFT), allowing both lattice parameters and atomic positions to fully relax prior to property evaluation.

All DFT calculations were carried out using Vienna Ab initio Simulation Package (VASP)~\cite{VASP1,VASP2,VASP3}. The generalized gradient approximation (GGA) \cite{GGA_PBE}, in the  Perdew-Burke-Ernzerhof (PBE) implementation, was employed as the exchange-correlation function with the projector augmented wave (PAW) method \cite{VASP3}. All structural relaxations were carried out using an electronic energy convergence criterion of 10$^{-7}$~eV, and the atomic positions were optimized until the residual forces on each atom were below 0.01~eV/{\AA}. The converged values for the energy cutoff is 800~eV for all three phases whereas k-point mesh are: 6 $\times$ 6 $\times$ 4 for the monoclinic phase, 6 $\times$ 6 $\times$ 6 for the orthorhombic phase, and 8 $\times$ 8 $\times$ 10 for the tetragonal phase. Structural relaxation was observed to introduce slight symmetry breaking. Since lattice dynamical calculations are particularly sensitive to such distortions in symmetry, the original, unrelaxed unit cells of the monoclinic and the tetragonal phase have been used for further calculations. The lattice parameters used in the present calculations are listed in Tab.~\ref{tab:UnitCellParams_WO3}.  

\begin{table}[bt]
    \caption{\label{tab:UnitCellParams_WO3} Unit cell parameters of the chosen polymorphs of WO$_3$ used in this study}
    \begin{ruledtabular}
        \centering
        
        \begin{tabular}{lccccccc}
            Phase & Space group (No.) & a (\AA) & b (\AA) & c (\AA) & $\alpha$ ($^\circ$) & $\beta$ ($^\circ$) & $\gamma$ ($^\circ$) \\
            \colrule
            Monoclinic & P2$_1$/c (14) & 5.306 & 5.207 & 7.693 & 90 & 91.5 & 90 \\
            Orthorhombic & Pbcn (60) & 7.500 & 7.869 & 7.712 & 90 & 90 & 90 \\
            Tetragonal & P4/nmm (129) & 5.325 & 5.325 & 3.944 & 90 & 90 & 90 \\
        \end{tabular}
    \end{ruledtabular}
\end{table}

The temperature-dependent effective potential (TDEP) method was utilised to obtain atomic forces and subsequently fit the interatomic force constants (IFCs) \cite{TDEP,TDEP_2FC,TDEP_3FC,TDEP_4FC}. Specifically, incorporating stochastic sampling within the TDEP approach (s-TDEP) \cite{TDEP_canonical_configuration1,TDEP_canonical_configuration2, sTDEP-1, ziqi_APL, aziz_PRB} efficiently captures finite-temperature anharmonic effects while avoiding a substantial increase in computational cost. 

Table~\ref{tab:IFC_Params} summarizes all parameters employed in the DFT calculations, including those used for stochastic sampling convergence, the extraction of finite-temperature IFCs, and phonon scattering calculations within the Boltzmann transport equation (BTE) framework. Furthermore, for all phases, the non-analytical term correction as well as isotopic scattering \cite{isotopic} were included to account for the effect of long-range dipole-dipole interaction and scattering due to naturally occurring isotopic concentrations in these crystals, respectively. 
\begin{table}[tb]
    \caption{\label{tab:IFC_Params} Converged parameters used for lattice dynamics calculations of each phase of WO$_3$, i.e., DFT calculations for estimation of forces using the s-TDEP method, subsequent determination of IFCs, and computation of scattering rates using BTE an implemented in FourPhonon.}
    \begin{ruledtabular}
        \centering
        
        \begin{tabular}{lccc}
            Paramater & Monoclinic & Orthorhombic & Tetragonal \\
            \colrule
            Temperature (K) & 400 & 700 & 1300 \\
            No. of atoms in unit cell & 16 & 32 & 8 \\
            Super cell size & 3~$\times$~3~$\times$~2 & 2~$\times$~2~$\times$~2 & 3~$\times$~3~$\times$~4 \\
            Energy cut-off (eV) & 800 & 800 & 800 \\
            k-point mesh & 2~$\times$~2~$\times$~2 & 3~$\times$~3~$\times$~3 & 2~$\times$~2~$\times$~2 \\ 
            Cutoff radius (2nd order IFC) (\AA) & max (8.0) & max (8.0) & max (8.0) \\ 
            Cutoff radius (3rd order IFC) (\AA) & 5.0 & 4.0 & 7.0 \\ 
            Cutoff radius (4th order IFC) (\AA) & 3.0 & 1.75 & 4.0 \\ 
            q-grid mesh (for BTE) & 10~$\times$~10~$\times$~10 & 10~$\times$~10~$\times$~10 & 10~$\times$~10~$\times$~10 \\
            
        \end{tabular}
    \end{ruledtabular}
\end{table}
Calculations for scattering rates was done using a modified version of the FourPhonon~\cite{FourPhonon} module using the sampling method \cite{Ziqi_sampling}, built on the ShengBTE~\cite{ShengBTE_2014} framework. Optical properties are further extracted using the added functionality in FourPhonon as described in our previous work \cite{SREE_P1_OPTICS}. Here, the crystal-radiation interaction is modelled as a resonant process where the incident photons couple with infrared-active transverse optical (TO) phonons close to the Brillouin zone center. The resulting dielectric response is formulated as:
\begin{equation}
    \label{eq:dieletric_response}
     \varepsilon_{\mu\nu}(\omega, T) = \varepsilon_\infty + \frac{4\pi}{\hbar V_{\rm{cell}}} \sum_{j} \frac{S_{\mu\nu}^j(T)}{\omega_{0, j}^2(T) - \omega^2 + 2\omega_{0,j}(T) \Sigma_j(\omega_{0,j},T)} 
\end{equation}
where $V_{cell}$ is the volume of the unit cell, $\varepsilon_\infty$ is the high-frequency dielectric constant, $\omega_{0,j}$ is the frequency of the $j$-th IR-active TO phonon mode, $S_{\mu\nu}^{j}$ is the oscillator strength, and 
 $\Sigma_j=\Delta_j-i\Gamma_j$ is the phonon self-energy,
where $\Delta_j$ is the anharmonic frequency shift and $\Gamma_j$ is the
phonon damping obtained from the calculated scattering rates. The calculated dielectric response thus arises solely from interactions between the incident electromagnetic field and infrared-active phonons. Free-carrier
contributions are neglected within the present formalism
\cite{SREE_P1_OPTICS}.

For each optical phonon mode, the calculated mode dipole moment is projected
onto directions parallel and perpendicular to the propagation direction of
the incident electromagnetic radiation, yielding the longitudinal and
transverse components, respectively. Since the electromagnetic field is
transverse, only phonon modes with a non-zero transverse dipole component
are infrared active and therefore contribute to the dielectric response.
The corresponding transverse projection determines the oscillator strength
$S_{\mu\nu}^{j}$ appearing in Eq.~(\ref{eq:dieletric_response}).
Temperature dependence is naturally incorporated through the TDEP-derived
phonon frequencies and self-energies, enabling the optical response to be
evaluated for each polymorph within its thermodynamically stable
temperature range.

\section{\label{sec:Results} Results and discussion}

\begin{figure*}[b]
\centering
\begin{minipage}{0.32\linewidth}
\includegraphics[width=\linewidth]{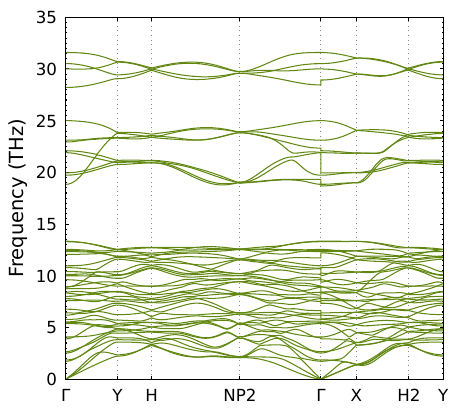}
\centerline{(a) Monoclinic - 400~K}
\end{minipage}
\begin{minipage}{0.32\linewidth}
\includegraphics[width=\linewidth]{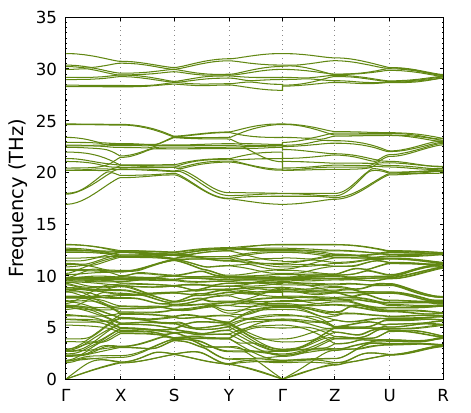}
\centerline{(b) Orthorhombic - 700~K}
\end{minipage}
\begin{minipage}{0.32\linewidth}
\includegraphics[width=\linewidth]{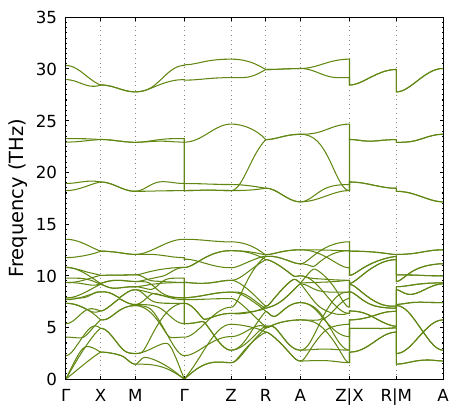}
\centerline{(c) Tetragonal - 1300~K}
\end{minipage}
\caption{\label{fig:Phonon_dispersions}
Phonon dispersions of three phases of WO$_3$ at their respective temperatures. For the monoclinic phase, ``NP2''
denotes an unnamed high-symmetry point along the selected path.}
\end{figure*}

Phonon dispersions for all three phases of tungsten trioxide are shown in Fig.~\ref{fig:Phonon_dispersions}. In contrast to several previous studies, no imaginary frequencies are observed, indicating the dynamical stability of each phase at its respective temperature. This result highlights the importance of incorporating temperature-induced renormalization effects via
the stochastically sampled TDEP method. 
To the best of our knowledge, no inelastic neutron scattering measurements have been reported for the phases considered here,  preventing a direct comparison with experiment. Likewise, previously reported first-principles studies are not directly comparable because they neglect the temperature-induced renormalization effects considered in the present work.

Analysis of the phonon eigenvectors shows that the tetragonal phase supports phonon modes that are purely TO and LO  with respect to the selected propagation direction. In
contrast, the monoclinic phase exhibits modes with mixed TO--LO character, consistent with its lower symmetry. Similar mixing is also observed in the orthorhombic phase, which we attribute to the slight symmetry breaking
introduced during structural relaxation. For modes with mixed TO--LO character, only the transverse component couples to the
incident electromagnetic radiation and therefore contributes to the
oscillator strength, $S_{\mu\nu}^{j}$, in Eq.~(\ref{eq:dieletric_response}).

\begin{figure*}[h]
\centering
\begin{minipage}{0.48\linewidth}
\includegraphics[width=\linewidth,page=1]{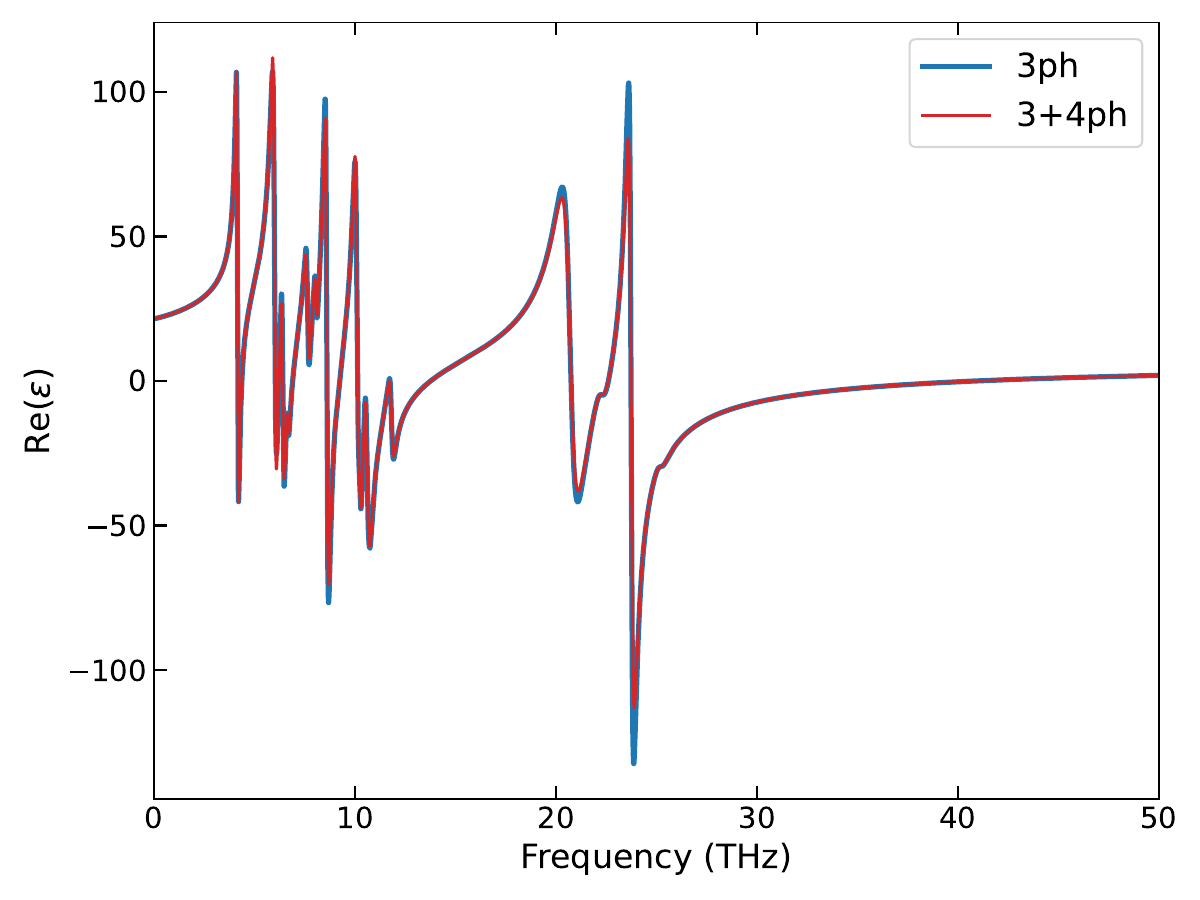}
\end{minipage}
\begin{minipage}{0.48\linewidth}
\includegraphics[width=\linewidth,page=2]{Images/Eps-WO3-MonC_400K.pdf}
\end{minipage} 
\centerline{(a) Monoclinic (400 K), propagation along the a-axis}

\begin{minipage}{0.48\linewidth}
\includegraphics[width=\linewidth,page=1]{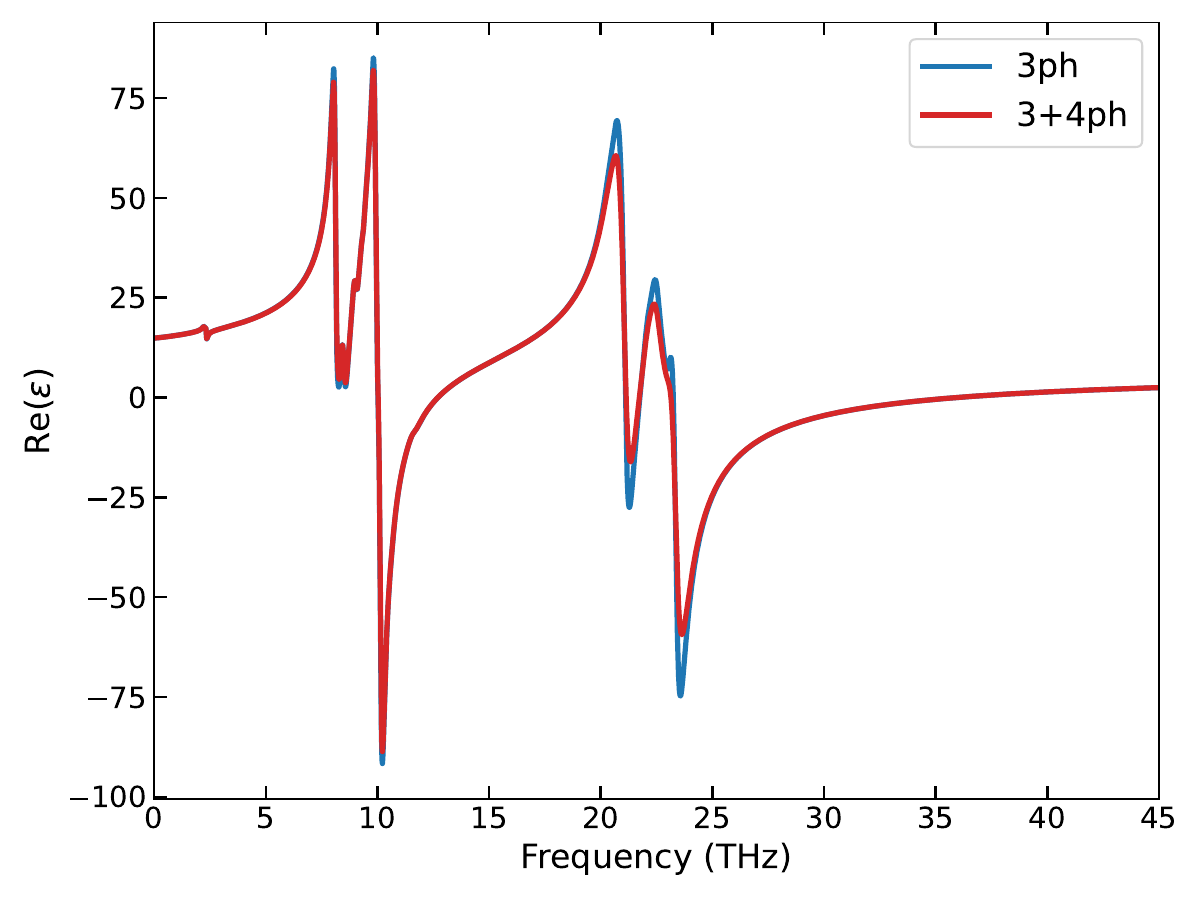}
\end{minipage}
\begin{minipage}{0.48\linewidth}
\includegraphics[width=\linewidth,page=2]{Images/Eps-WO3-OrtR_700K.pdf}
\end{minipage} 
\centerline{(b) Orthorhombic (700~K), propagation along the a-axis}

\begin{minipage}{0.48\linewidth}
\includegraphics[width=\linewidth,page=1]{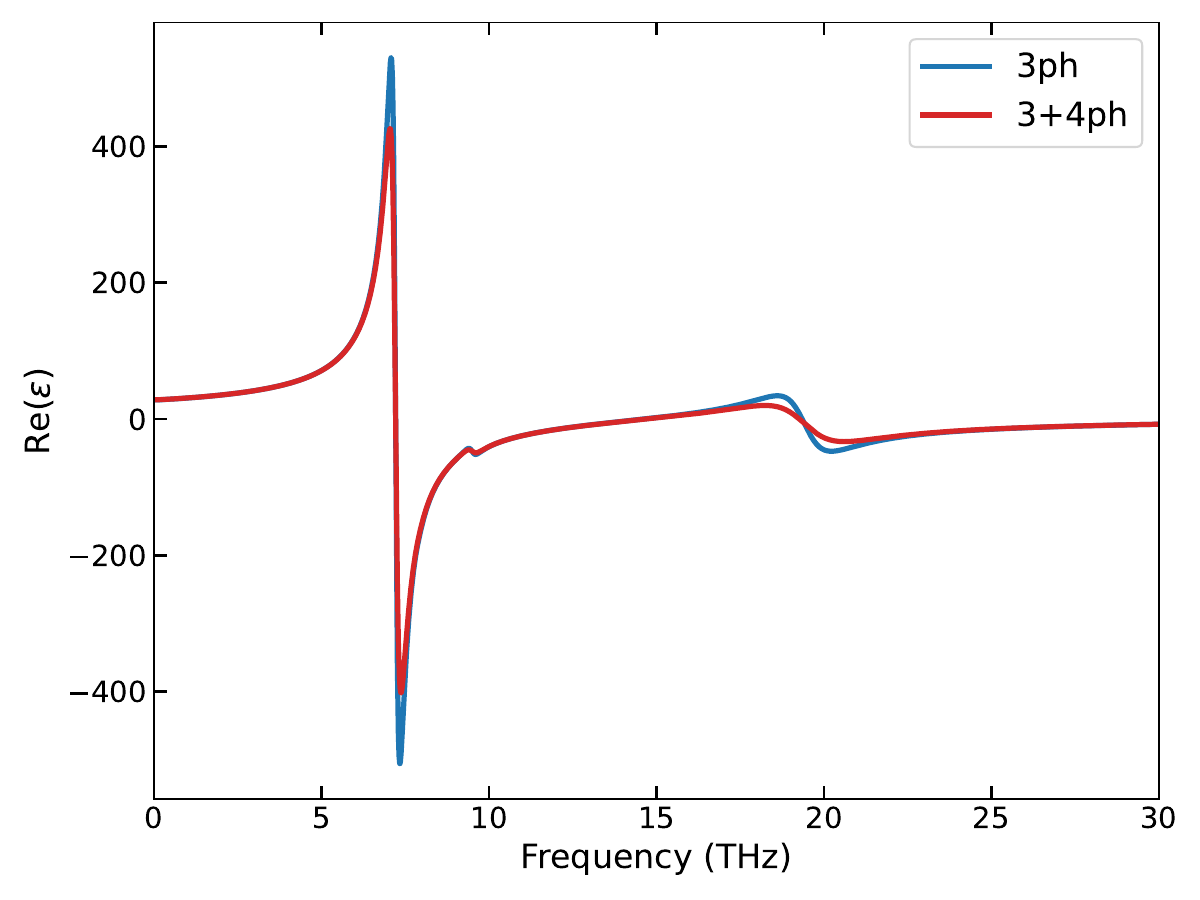}
\end{minipage}
\begin{minipage}{0.48\linewidth}
\includegraphics[width=\linewidth,page=2]{Images/Eps-WO3-TetG_1300K.pdf}
\end{minipage} 
\centerline{(c) Tetragonal (1300~K), propagation along the c-axis}

\caption{\label{fig:Eps}
Real (left) and imaginary (right) parts of the dielectric function,
$\varepsilon$, of the monoclinic (400~K), orthorhombic (700~K), and
tetragonal (1300~K) phases of WO$_3$. The dielectric response is evaluated
for the indicated direction of propagation of the incident radiation.
Results obtained using only three-phonon scattering (blue) are compared
with those including both three- and four-phonon scattering (red).}  
\end{figure*}

Figure~\ref{fig:Eps} presents the real and imaginary parts of the dielectric
function for the three crystallographic phases of WO$_3$, evaluated for the
chosen direction of propagation of the incident radiation. 
All three phases exhibit anisotropic optical behaviour owing to their
non-cubic crystal symmetry.
The optical response along the other crystallographic directions is presented in Fig.~\ref{fig:aniso_MonC},~\ref{fig:aniso_OrtR} and~\ref{fig:aniso_TetG} in Appendix~\ref{appx:Anisotropic_ops}. In addition,
the frequency-dependent dielectric function and refractive index for all
three phases are provided in Ref.~\onlinecite{Op_Const_Data_Github}.
A comparison of the three-phonon and combined three- plus four-phonon
results in Fig.~\ref{fig:Eps} shows that fourth-order phonon--phonon
scattering has the greatest influence on the tetragonal phase, producing
the largest changes in both the resonance linewidths and peak intensities.
In contrast, the monoclinic phase exhibits comparatively weak anharmonicity,
while in the orthorhombic phase the effect of fourth-order scattering is
primarily confined to the high-frequency optical modes between
20 and 25~THz.

\begin{figure*}[h]
\centering
\begin{minipage}{0.48\linewidth}
\includegraphics[width=\linewidth]{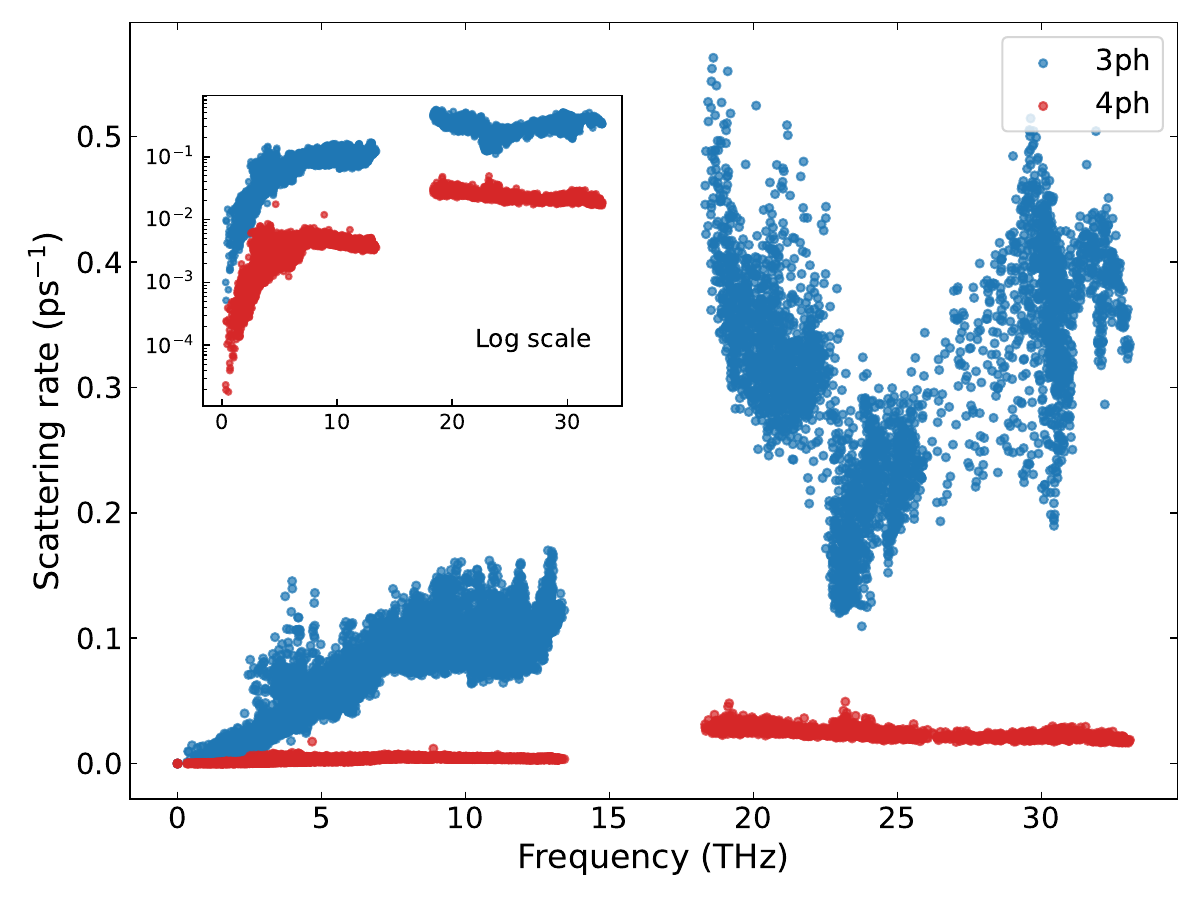}
\centerline{(a) Monoclinic (400~K)}
\end{minipage} 
\\
\begin{minipage}{0.48\linewidth}
\includegraphics[width=\linewidth]{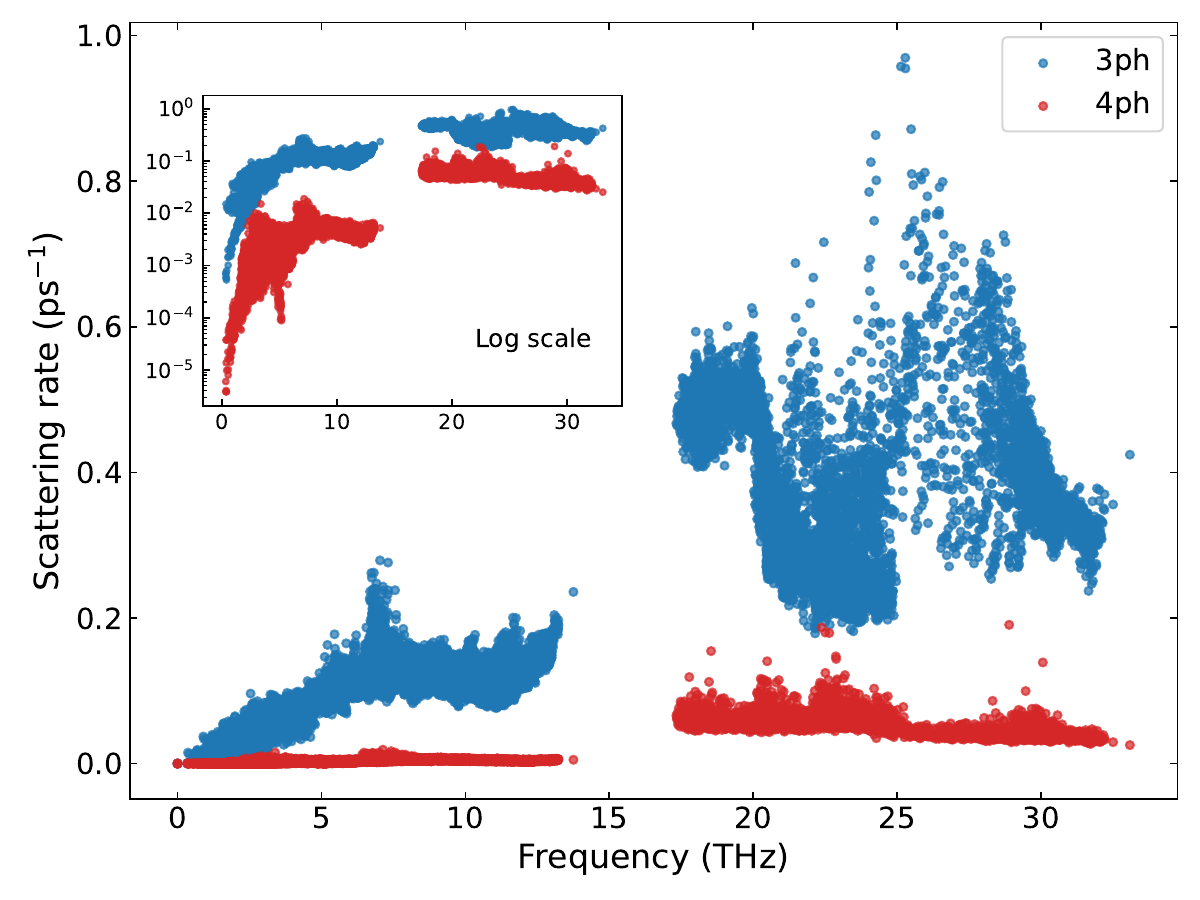}
\centerline{(b) Orthorhombic (700~K)}
\end{minipage}
\\
\begin{minipage}{0.48\linewidth}
\includegraphics[width=\linewidth]{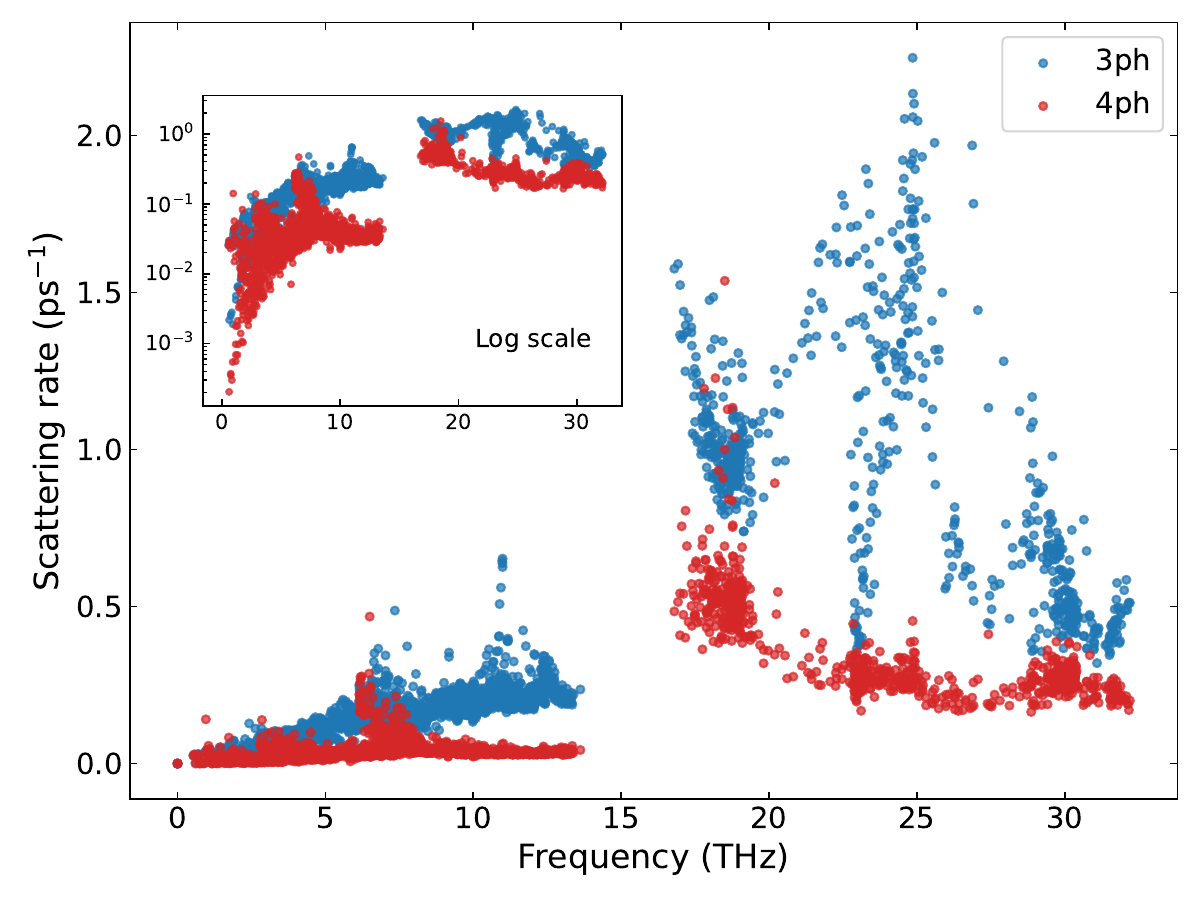}
\centerline{(c) Tetragonal (1300~K)}
\end{minipage}
\caption{\label{fig:Scatt_rates}
Comparison of three-phonon (blue) and four-phonon (red) scattering rates
as a function of phonon frequency for the three crystallographic phases of
WO$_3$ at their respective temperatures. 
The inset in each panel shows the same data on a logarithmic scale to highlight the relative magnitudes of the three- and four-phonon
scattering rates.}
\end{figure*}

This trend is further corroborated by the calculated phonon scattering
rates (Fig.~\ref{fig:Scatt_rates}) and lattice thermal conductivities
(Table~\ref{tab:therm_props}). Among the three polymorphs, the tetragonal
phase exhibits the largest increase in scattering rates upon inclusion of
four-phonon processes, particularly for the optical phonon branches,
consistent with its stronger anharmonic character inferred from the
dielectric spectra. Since the same anharmonic interactions also govern
phonon-mediated thermal transport, the lattice thermal conductivity
provides an independent measure of the relative anharmonicity of the three phases. As listed in Table~\ref{tab:therm_props}, inclusion of four-phonon scattering reduces the thermal conductivity by nearly 20\% for the tetragonal phase, whereas the reduction is less than 6\% for the monoclinic and orthorhombic phases. Since no experimental thermal conductivity data are currently available for comparison, we instead assess the reliability of the underlying lattice-dynamical calculations by comparing the calculated constant-volume heat capacity, $C_v$, with the experimentally measured constant-pressure heat capacity, $C_p$, reported in Ref.~\onlinecite{sp_heat_cap_lit}. As the simulated temperatures are well removed from the phase-transition temperatures, the approximation $C_p \approx C_v$ is appropriate \cite{cp_cv_Callen}. The good agreement shown in Table~\ref{tab:therm_props} provides further
validation of the lattice-dynamical framework used to predict the infrared
optical properties presented in this work.

\begin{table}[tb]
    \caption{\label{tab:therm_props}     
    Calculated thermal conductivity of the WO$_3$ phases considering three-phonon (3Ph) and combined three- and four-phonon (3+4Ph) scattering. 
    The non-analytical correction at the $\Gamma$-point was evaluated by approaching $\Gamma$ along the crystallographic a-axis for all phases.
    The table also compares the calculated $C_v$ with the experimentally measured $C_p$ from literature 
    }
    \begin{ruledtabular}
        \centering
        
        \begin{tabular}{lccc}
             & Monoclinic & Orthorhombic & Tetragonal \\
            \colrule
            T (K) & 400 & 700 & 1300 \\
            $\kappa_{3Ph}$ (W~m$^{-1}$~K$^{-1}$)& 6.26 & 3.71 & 5.06 \\  
            $\kappa_{3+4Ph}$ (W~m$^{-1}$~K$^{-1}$)& 5.96 & 3.58 & 4.09 \\ 
            C$_v$ (J~mol$^{-1}$~K$^{-1}$) & 81.6 & 92.6 & 97.7 \\
            C$_p$ (J~mol$^{-1}$~K$^{-1}$) -- Ref.~\cite{sp_heat_cap_lit} & 83.25 & 95.17 & 100.6 \\
        \end{tabular}
    \end{ruledtabular}
\end{table}

We have thus, within the assumptions of the present framework, characterized the infrared optical properties of the three crystallographic phases of WO$_3$. Although direct experimental validation of the predicted optical properties is presently unavailable, the substantial phase-dependent differences predicted here demonstrate
the strong influence of crystal symmetry and lattice dynamics on the infrared response, and provide a basis for future experimental and theoretical investigations.

More broadly, this work demonstrates that, given accurate crystallographic information, the infrared optical response of polymorphic materials can be systematically predicted from first principles within a unified lattice-dynamical framework. The substantial differences observed among the three WO$_3$ polymorphs illustrate how structural phase transformations can significantly modify the infrared optical response through changes in the underlying lattice dynamics. The methodology is readily extendable to other polymorphic materials with well-characterized crystal structures, enabling predictive evaluation of their phase-dependent infrared optical properties
over their respective stability ranges.


\section{\label{sec:Summary} Summary}

In summary, we have predicted the frequency-dependent dielectric function
of three temperature-dependent crystallographic phases of tungsten trioxide
(WO$_3$) from first principles. By incorporating temperature-renormalized lattice dynamics within the TDEP framework, the infrared optical properties of each polymorph are evaluated within its thermodynamically stable temperature regime. The results demonstrate that temperature-induced structural phase transformations produce pronounced changes in the infrared optical response through modifications of the crystal symmetry, lattice dynamics, and anharmonic phonon interactions. Among the three phases considered, the tetragonal phase exhibits the strongest anharmonic character, resulting in the largest influence of four-phonon scattering on the predicted optical properties. More broadly, this work establishes a first-principles framework for systematically investigating phase-dependent infrared optical properties in polymorphic materials, thereby enabling computational discovery and design of materials with tunable thermochromic behaviour.

\begin{acknowledgments}
Simulations were performed at the Rosen Center for Advanced Computing (RCAC) of Purdue University.

S.S. acknowledges financial support from SERB for his award [SB/S9/Z-03/2017-XXIII (2022)] under the Overseas Visiting Doctoral Fellowship (OVDF) program at Purdue University, West Lafayette, USA. 
S.S. also acknowledges support from Prime Minister’s Research Fellowship (PMRF). 
\end{acknowledgments}

\appendix

\section{Anisotropic properties}
\label{appx:Anisotropic_ops}

\begin{figure*}[h]
\centering

\begin{minipage}{0.45\linewidth}
\includegraphics[width=\linewidth,page=1]{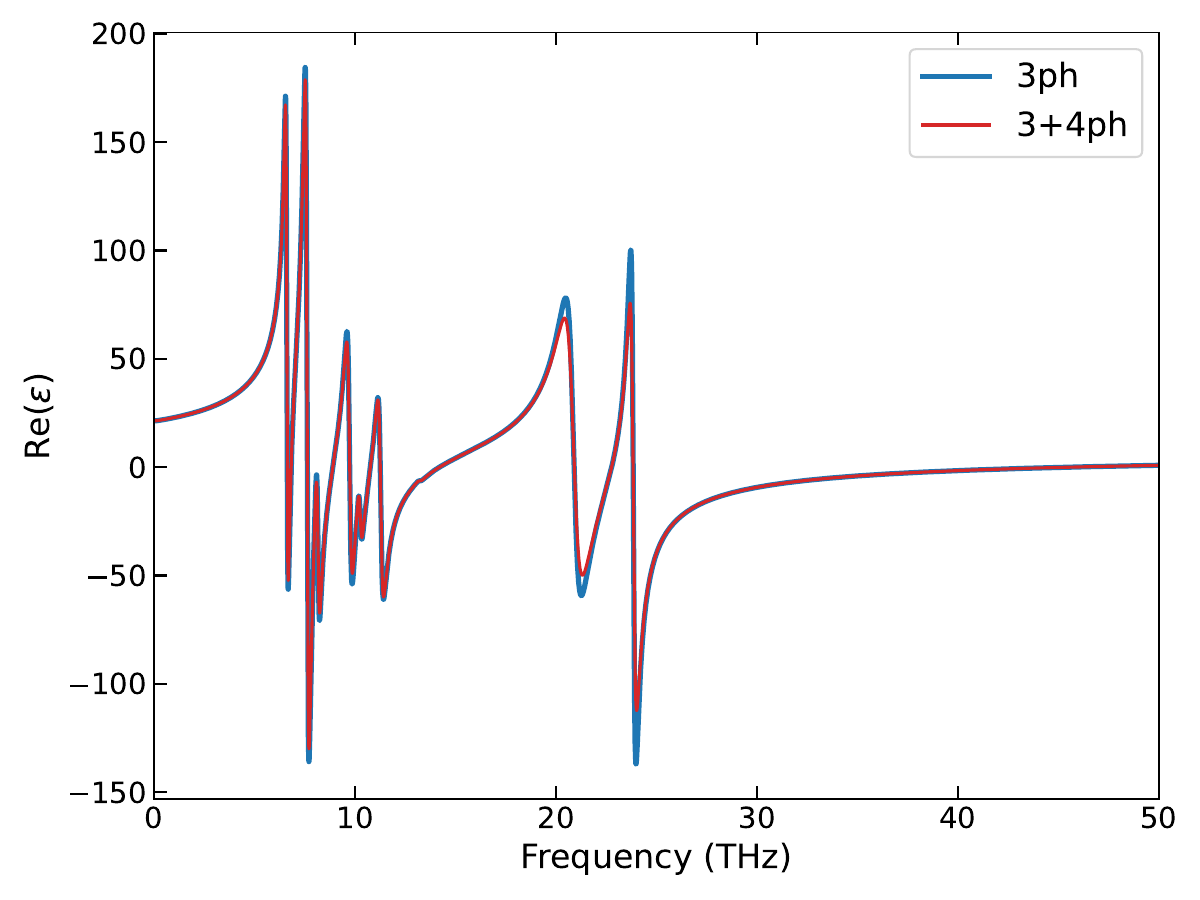}
\end{minipage}
\begin{minipage}{0.45\linewidth}
\includegraphics[width=\linewidth,page=2]{Images/anisotropicProps/Eps-WO3-MonC_010.pdf}
\end{minipage} 
\centerline{(a)}


\begin{minipage}{0.45\linewidth}
\includegraphics[width=\linewidth,page=1]{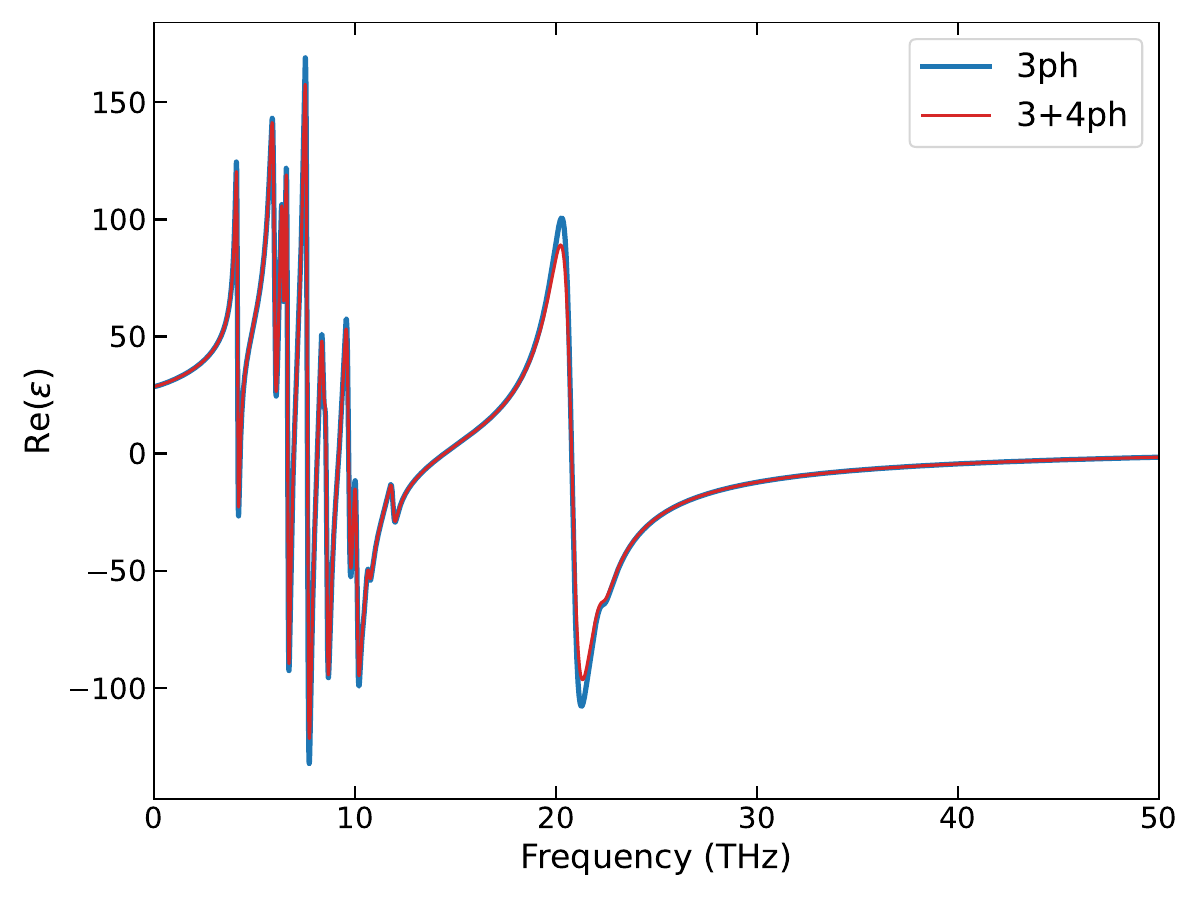}
\end{minipage}
\begin{minipage}{0.45\linewidth}
\includegraphics[width=\linewidth,page=2]{Images/anisotropicProps/Eps-WO3-MonC_001.pdf}
\end{minipage} 
\centerline{(b)}


\caption{\label{fig:aniso_MonC}
Real (left) and imaginary (right) parts of the dielectric function $\varepsilon$ 
of the monoclinic phase of WO$_3$ at 400~K, when propagation direction of the incident radiation is along (a) the b-axis and (b) the c-axis, showcasing its anisotropic nature. Both third- (blue solid line) and fourth-order (red solid line) scattering has been included.}  

\end{figure*}

\begin{figure*}[h]
\centering

\begin{minipage}{0.45\linewidth}
\includegraphics[width=\linewidth,page=1]{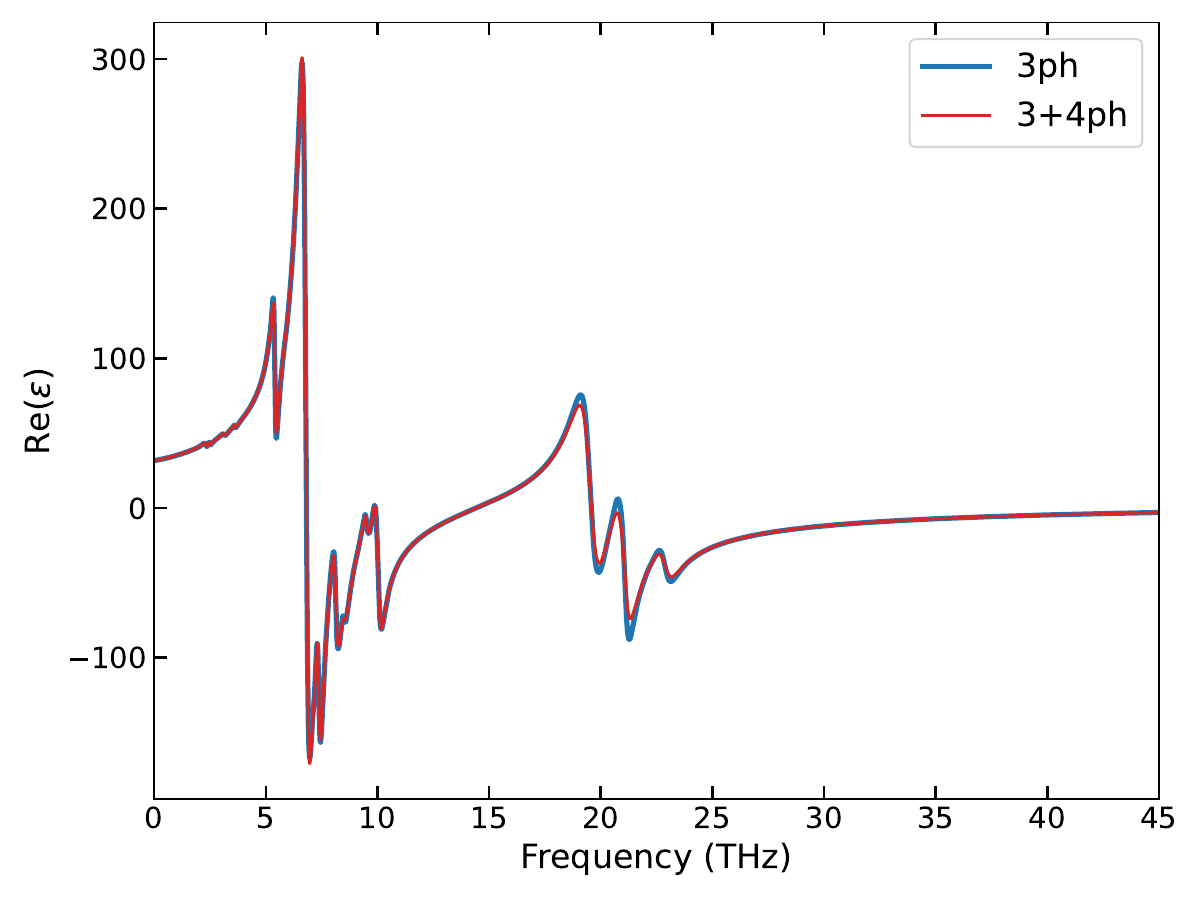}
\end{minipage}
\begin{minipage}{0.45\linewidth}
\includegraphics[width=\linewidth,page=2]{Images/anisotropicProps/Eps-WO3-OrtR_010.pdf}
\end{minipage} 
\centerline{(a)}


\begin{minipage}{0.45\linewidth}
\includegraphics[width=\linewidth,page=1]{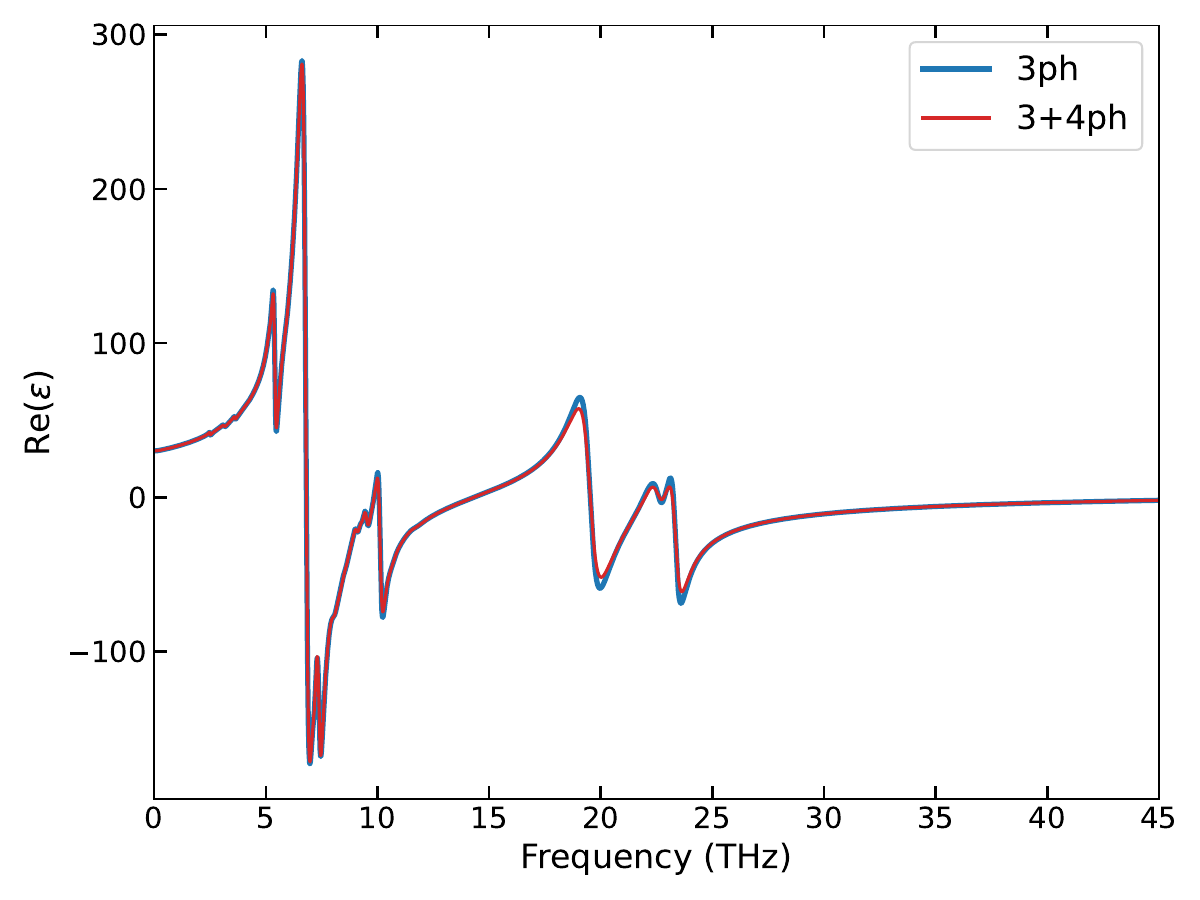}
\end{minipage}
\begin{minipage}{0.45\linewidth}
\includegraphics[width=\linewidth,page=2]{Images/anisotropicProps/Eps-WO3-OrtR_001.pdf}
\end{minipage} 
\centerline{(b)}


\caption{\label{fig:aniso_OrtR}
Real (left) and imaginary (right) parts of the dielectric function $\varepsilon$ 
of the orthorhombic phase of WO$_3$ at 700~K, when propagation direction of the incident radiation is along (a) the b-axis and (b) the c-axis, showcasing its anisotropic nature. Both third- (blue solid line) and fourth-order (red solid line) scattering has been included.}  

\end{figure*}

\begin{figure*}[h]
\centering

\begin{minipage}{0.45\linewidth}
\includegraphics[width=\linewidth,page=1]{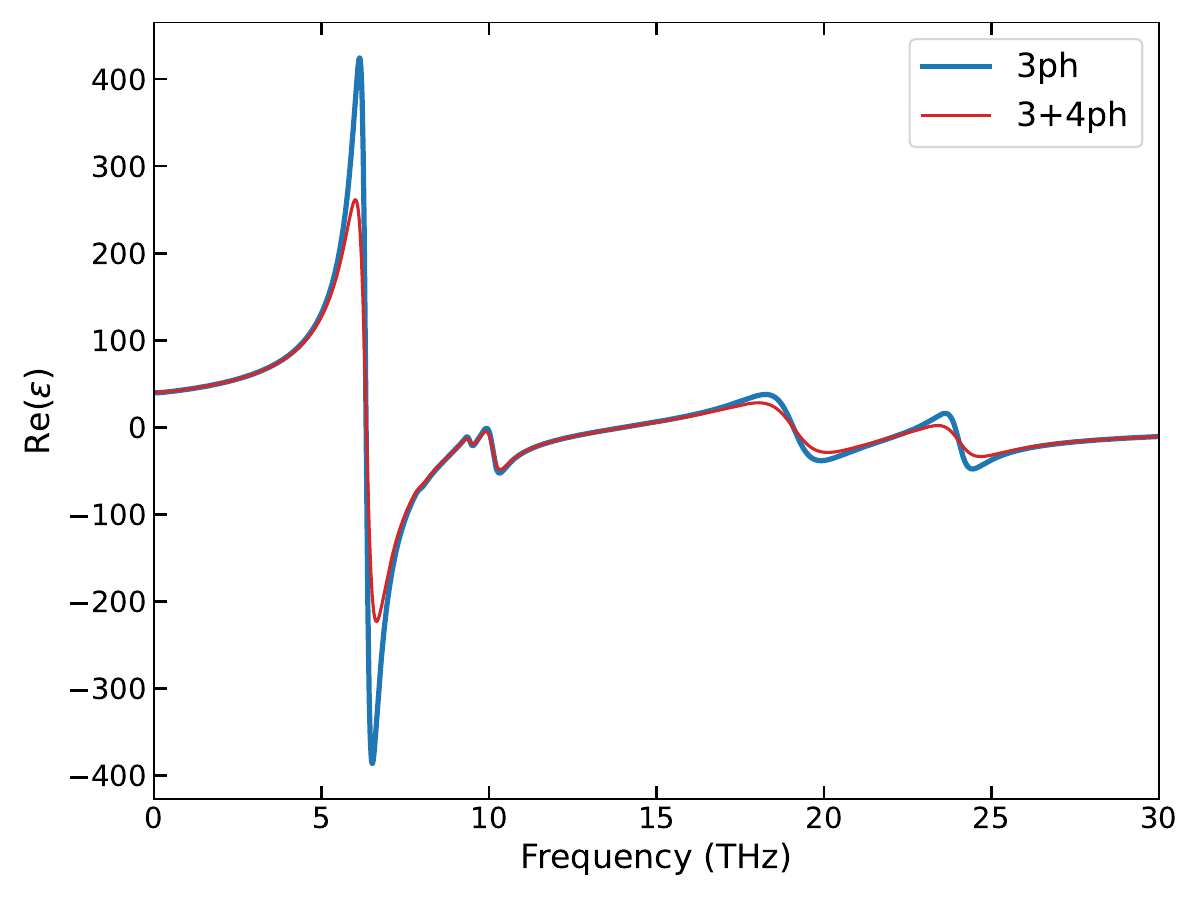}
\end{minipage}
\begin{minipage}{0.45\linewidth}
\includegraphics[width=\linewidth,page=2]{Images/anisotropicProps/Eps-WO3-TetG_100.pdf}
\end{minipage} 


\caption{\label{fig:aniso_TetG}
Real (left) and imaginary (right) parts of the dielectric function $\varepsilon$ of the tetragonal phase of WO$_3$ at 1300~K when direction of propagation of incident radiation is along the a-/b- axis. Both third- (blue solid line) and fourth-order (red solid line) scattering has been included.}  

\end{figure*}

Figures~\ref{fig:aniso_MonC},~\ref{fig:aniso_OrtR} and~\ref{fig:aniso_TetG} illustrate the anisotropic behaviour exhibited by the three phases of WO$_3$ vis-a-vis the calculated infrared optical properties.

\bibliography{apssamp}

\end{document}